\documentclass[prd,twocolumn,showpacs,preprintnumbers,nofootinbib,amsmath,amssymb,nobalancelastpage]{revtex4}
\usepackage[pdftex]{graphicx}
\usepackage{latexsym,amsmath,amssymb,lmodern,float,url}
\usepackage{natbib}  
\usepackage[pdftex,bookmarks,linktocpage,pdfpagelabels,plainpages=false,hyperfigures,linkcolor=blue,citecolor=blue,urlcolor=blue]{hyperref} \usepackage{xspace}
\usepackage{courier}
\hypersetup{colorlinks=true}
\usepackage{bm}
\usepackage{dcolumn}
\usepackage{amsmath}
\usepackage{amssymb}
\usepackage{color}
\usepackage{xcolor}
\usepackage{soul}
\usepackage{url}
\usepackage{float}

\usepackage{aas_macros}
\usepackage[normalem]{ulem}
\usepackage{comment}

\def\Fig#1{Fig.~(\ref{#1})}
\def\Sec#1{Sec.~\ref{#1}}
\def\Tab#1{Tab.~\ref{#1}}

\def\Mpc{\rm{Mpc}}

\def\lcdm{\Lambda \mathrm{CDM}}
\def\H0{$H_0$}
\def\si8{$\sigma_8$}
\def\S8{$S_8$}
\def\OmegaM{\Omega_{\mathrm{m}}}

\graphicspath{{./figs/}}

\begin{document}

\title{The $H_0$ and $S_8$ tensions necessitate early and late time changes to $\Lambda$CDM}

\author{Steven~J.~Clark} 
\email{steven$\_$j$\_$clark@brown.edu}
\affiliation{ Department of Physics, Brown University, Providence, RI 02912-1843, USA}
\affiliation{ Brown Theoretical Physics Center, Brown University, Providence, RI 02912-1843, USA}

\author{Kyriakos Vattis}
\email{kyriakos$\_$vattis@brown.edu}
\affiliation{ Department of Physics, Brown University, Providence, RI 02912-1843, USA}
\affiliation{ Brown Theoretical Physics Center, Brown University, Providence, RI 02912-1843, USA}

\author{JiJi Fan}
\email{jiji$\_$fan@brown.edu}
\affiliation{ Department of Physics, Brown University, Providence, RI 02912-1843, USA}
\affiliation{ Brown Theoretical Physics Center, Brown University, Providence, RI 02912-1843, USA}

\author{Savvas M. Koushiappas}
\email{koushiappas@brown.edu}
\affiliation{ Department of Physics, Brown University, Providence, RI 02912-1843, USA}
\affiliation{ Brown Theoretical Physics Center, Brown University, Providence, RI 02912-1843, USA}

\date{\today}

\begin{abstract}
An only early or only late time alteration to $\lcdm$ has been inadequate at resolving both the \H0 and \S8 tensions simultaneously; however, a combination of early and late time alterations to $\lcdm$ can provide a solution to both tensions. As an illustration, we examine a combined Early Dark Energy - Decaying Dark Matter model. While early dark energy has the ability to resolve the \H0 tension, it leads to a discrepancy in \S8 measurements. We show that the addition of decaying dark matter helps resolve the \S8 discrepancy that would otherwise be enhanced in an early dark energy model, while the latter is able to relieve the \H0 disagreement to within the 95th percentile interval. Our results show a preference for the combined model over $\lcdm$ with {$\Delta \rm{AIC} = -6.72$}, hinting that both early and late universe modifications may be necessary to address the cosmological tensions.
\end{abstract}

\maketitle

\section{Introduction}

The standard cosmological model known as $\lcdm$, consisting mostly of dark energy (in the form of a cosmological constant $\Lambda$) and Cold Dark Matter (CDM) has come under increasing scrutiny in recent years. The emergence of tensions between the value of the present-day Hubble parameter \H0 as inferred from early time cosmology using Cosmic Microwave Background (CMB) measurements \cite{2020A&A...641A...6P} and that from local late time cosmology using Type Ia Supernovae \cite{2016ApJ...826...56R, 2018ApJ...861..126R, 2019arXiv190307603R} has been considered extensively in the literature as possible evidence for the existence of new physics beyond the standard cosmology scenario. A similar discrepancy appears in the amplitude of the variance of the matter density field on scales of $8 h^{-1} {\mathrm{Mpc}}$,  $\sigma_8$ (or equivalently in  \S8 = \si8$(\OmegaM/0.3)^{0.5}$, where $\OmegaM$ is the total matter density of the universe).  Similar to the \H0 tension, late universe measurements of \S8 are in an apparent disagreement with the value of \S8 inferred from the CMB \cite{2009ApJ...692.1060V,2013PhRvL.111p1301M,2015PhRvD..91j3508B,2015MNRAS.451.2877M,2016PhRvD..93d3522R,2020A&A...641A...6P,2020arXiv200211124D,2018A&A...614A..13S,2019arXiv190105289D}. Both of these tensions could be potential evidence that $\lcdm$ does not fully describe the observable universe.

Unaccounted systematic uncertainties, in particular, those with Cepheids and supernovae, have been proposed as causes for the \H0 tension \cite{2018arXiv181002595S,2018arXiv181003526R,2018arXiv181004966B}. But this class of explanation appears to be less favorable with a variety of new data sets coming from various sources \cite{2021arXiv210301183D} confirming the tension at the 4.4 - 6$\sigma$ level. More recently it has been proposed that systematic uncertainties related to the choice of Cepheid color-luminosity calibration method \cite{Mortsell:2021nzg} could affect the ability of the distance ladder to measure the value of \H0 to the required precision. So far a large array of both early and late universe modifications to $\lcdm$ have been proposed: they are summarized in two recent thorough reviews \cite{2021arXiv210301183D, 2021arXiv210710291S}. While there has been no clear preferred solution to date, the work of Knox \& Millea \cite{2020PhRvD.101d3533K} points out that early universe solutions are ``less unlikely." However, it seems that early universe solutions could unfortunately fail to agree with large scale structure observations as shown in \cite{2019Symm...11.1035V,2020PhRvD.102d3507H,2020PhRvD.102j3502I,2020arXiv201004158J} and make the \S8 tension even more prominent.

The \S8 tension, while not as statistically significant (ranging between 1.5 - 2.5$\sigma$) has  received a lot of attention as well. Most proposed solutions introduce some form of self interactions in the dark sector \cite{2016PhRvD..94d3518P, 2014PhRvL.113r1301S, 2014PhRvD..89h3517Y,2018arXiv181202333V, 2017EL....12039001G,2018MNRAS.478..126G,2017PhLB..768...12K,2016PhLB..762..462K} in an attempt to erase structure in the late universe. Other proposals include, but are not limited to, dark matter-neutrino interactions \cite{2018PhRvD..97d3513D}, modifications to gravity \cite{2018PhRvD..97j3503K}, or neutrino self-interactions \cite{2019arXiv190200534K}. The apparent correlation between the two tensions indicates that one tension cannot be addressed without the other.

One notable early universe solution for the \H0 tension is Early Dark Energy (EDE), an early period of dark energy domination that reduces the size of the acoustic horizon and thus increases the value of \H0 inferred from CMB measurements \cite{2019PhRvL.122v1301P,Agrawal:2019lmo,Lin:2019qug, 2019arXiv191010739N,2019arXiv191111760S}. To achieve this, the model introduces a scalar field that behaves like a cosmological constant at high redshifts ($z>3000$) and then gets diluted at the same rate as radiation or faster as the universe expands.\footnote{Implications of the ACT data set for EDE can be found in two recent works \cite{2021arXiv210904451H,2021arXiv210906229P}.} Unfortunately, a good fit of the model to the CMB power spectra requires a higher value of matter density at recombination than $\lcdm$ \cite{2020PhRvD.102d3507H,2020PhRvD.102j3502I}. This enhances structure formation at late times and increases the value of \S8. 

On the other hand, the introduction of Decaying Dark Matter (DDM) has been investigated as a candidate to solve both tensions simultaneously \cite{2021PhRvD.103d3014C,2019PhRvD..99l1302V, 2015PhRvD..92f1301A,2017JCAP...10..028B,2018PhRvD..98b3543B,Pandey:2019plg,1984MNRAS.211..277D,1985PAZh...11..563D,1988SvA....32..127D,Hooper:2011aj}. More specifically, late time decays of a massive cold parent particle decaying to one massless and one massive daughter particle of the form $\psi \rightarrow \gamma^\prime + \chi$ were discussed in \cite{2014PhRvD..90j3527B,2019PhRvD..99l1302V, 2021PhRvD.103d3014C, 2020arXiv200809615A,2021arXiv210212498A}. While the model looked promising initially, it was shown later that these decays cannot resolve the \H0 tension due to imprints induced at late times on low multipoles of the CMB power spectrum, which severely constrain the model. In addition, it was shown that decaying dark matter cannot relieve the \S8 tension when only early universe measurements are considered. But in a joint analysis with late time constraints on \S8, DDM shows a potential in alleviating the \S8 tension \cite{2021arXiv210212498A}. 

Despite all the efforts invested, it appears that {\it a single modification of $\lcdm$ in either the early or late universe has yet been successful in solving both tensions at the same time}. One example demonstrating the necessity for a binary modification to $\lcdm$, is a combination of EDE and additional ultra-light axion oscillating at early times ($z>10^4$) that suppresses the matter power spectrum  \cite{2021arXiv210412798A}. This dual modification reduces the tensions to $1.4 \sigma$ for \H0 and $1.2 \sigma$ for \S8. Another proposal includes two axions oscillating at different times to relax both tensions~\cite{2021JCAP...08..057F,Fung:2021fcj}. In a similar spirit, we examine the simultaneous effects of EDE and late-time DDM on the cosmological evolution in this article. We will show that the increase of the energy density at recombination due to an EDE component and the ability of DDM to erase the excess matter at later times could help relieve both \H0 and \S8 tensions. The paper is structured as follows: in \Sec{sec:Models}, we review the formalism of both components in our model and their cosmological implications, specifically on the CMB power spectra. In \Sec{sec:results} we present the results of a Markov Chain Monte Carlo analysis applied on a combination of CMB, Baryon Acoustic Oscillation (BAO), and Type Ia supernovae data. Finally we conclude in \Sec{sec:conclusion}.

%%%%%%%

\section{Model overview}\label{sec:Models}

The model we study in this work consists of two modifications to the standard cosmological model. The first one is a period of EDE as was proposed in \cite{2019PhRvL.122v1301P}; the second is the addition of DDM during late cosmological times as described in \cite{2014PhRvD..90j3527B,2016PhRvD..93b3510B}. In this section, we will give a brief summary of these two modifications.

\subsection{First component: Early Dark Energy} \label{sec:EDE}
The original EDE model utilizes a scalar field $\phi$ with a potential of $V(\phi)\propto [1-\cos (\phi/f) ]^n $, where $f$ is the field range of $\phi$ and $n$ indicates the power.\footnote{This potential is simply a convenient parametrization and we will not discuss its UV completions here.} To simplify the effects of EDE, we implement a fluid approximation of the system based on \cite{2018PhRvD..98h3525P,2019PhRvL.122v1301P}. 

However, in our model, we cannot express the EDE density as a function of redshift as easily as in e.g., \cite{2018PhRvD..98h3525P}, because the subsequent dark matter decays (see Sec.~\ref{sec:decaying_dm_mas}) alter the expansion history in a non-trivial way. 
Instead, we follow an alternative approach in which we define a simple characterization of the equation of state $w_{\rm EDE}(z)$ and introduce the parameter $\rho_{\rm EDE}(z=0)$, defined as the EDE energy density today. We also define the density parameter $\Omega_\mathrm{EDE}=\rho_{\rm EDE}(0)/\rho_{c,0}$, where $\rho_{c,0}$ is the critical energy density today. For $w_{\rm EDE}(z)$, we use a more general form of the equation of state from \cite{2020arXiv200809098B}, which can be seen as a generalized formulation of the CPL parameterization
\cite{2001IJMPD..10..213C, 2003PhRvL..90i1301L} 
commonly used to study dynamical dark energy in the late universe, 
\begin{equation}
w_{\rm EDE} = w_0 + \frac{w_a}{2} \left\{1-\tanh \left[ \alpha \log_{10}(a_{\rm EDE}/a) \right] \right\} \, . 
\end{equation}
Here, $w_0$ is the equation of state in the early universe before the EDE component oscillates. $w_a$ controls the change to the equation of state $w_{\rm EDE}$ after the field begins to oscillate such that at late times $w_\mathrm{EDE} = w_0+w_a=(n-1)/(n+1)$, where the second equality connects the equation of state to the power of the scalar potential $n$. The scale factor is as usual $a=(1+z)^{-1}$, and the midpoint of the equation of state's transition between early and late universe values ($w_\mathrm{EDE} = w_0 + w_a/2$) occurs at $a_\mathrm{EDE}$. $\alpha$ is a parameter that controls the rate of the transition period. We have set $\alpha$ to 5, in close agreement with the transition rate for the energy density in \cite{2018PhRvD..98h3525P}.\footnote{Connections between EDE cosmological quantities and $n$, the power in the scalar potential, could be found in \cite{2018PhRvD..98h3525P}.}

For the EDE perturbations, we follow \cite{2018PhRvD..98h3525P} and write the perturbation equations for $a>a_{\rm EDE}$ in the synchronous gauge as
\begin{align}
\dot{\delta}_{\rm EDE} = & -(1+w_{\rm EDE})\left( \theta_{\rm EDE} +\frac{\dot{h}}{2} \right) \nonumber \\
&- 3 (c_s^2 - w_{\rm EDE}) \mathcal{H} \delta_{\rm EDE} \nonumber \\
&- 9(1+w_{\rm EDE})(c_s^2-c_a^2)\mathcal{H} \frac{\theta_{\rm EDE}}{k^2} \, , \\
\dot{\theta}_{\rm EDE} = & -(1-3c_s^2)\mathcal{H} \theta_{\rm EDE} + \frac{c_s^2 k^2}{1 + w_{\rm EDE}} \delta_{\rm EDE} \, , 
\end{align}
where derivatives are with respect to conformal time, $h$ is the trace of the metric perturbation, $\mathcal{H}$ is the conformal Hubble expansion rate, $k$ is the wavenumber, $c_s^2$ is the effective sound speed, and $c_a^2$ is the adiabatic sound speed in the synchronous gauge. $c_a^2$ and $c_s^2$ are given by
\begin{align}
c_a^2 = & -\frac{3n+1}{n+1} = -2-w_n \, ,\\
c_s^2 = & \frac{2a^2(n-1)\overline{\omega}^2+k^2}{2a^2(n+1)\overline{\omega}^2+k^2} = \frac{4a^2 w_n \overline{\omega}^2+(1-w_n)k^2}{4a^2 \overline{\omega}^2+(1-w_n)k^2} \, .
\end{align}
Here, $w_n=(n-1)/(n+1)$ and $\overline{\omega}$ is the angular frequency of the oscillating field $\overline{\omega} = \overline{\omega}_0 a^{-3w_n} $. We choose parameters which give an enhanced \H0 in \cite{2019PhRvL.122v1301P}, namely $n=3$ and $\overline{\omega}_0 = 2 \times 10^{-4} \; \Mpc^{-1}$. For numerical purposes in the subsequent calculations, we set $w_0 = -0.9999$ and $w_a = 1.4999$.

\subsection{Second component: Decaying Dark Matter} \label{sec:decaying_dm_mas}
 We consider a massive cold parent particle decaying to one massless and one massive daughter particle.  Such models arise in extensions to the Standard Model that include Super WIMPs or excited dark fermions with magnetic dipole transitions \cite{Feng:2003xh, Choquette:2016xsw} (for cosmological implications from such models see \cite{2014PhRvD..90j3527B, 2016PhRvD..93b3510B, 2019PhRvD..99l1302V, 2021PhRvD.103d3014C}). 
 
 We denote such decays as $\psi \rightarrow \gamma^\prime + \chi$. From here on, we will label quantities of the particles involved using the subscripts $0$, $1$, and $2$ respectively.  The model also introduces two new parameters: the decay width $\Gamma$ and the fraction of the rest mass energy of the parent particle that is transferred to the massless daughter, $\epsilon$. 

The background density evolution of each species can be described as \cite{2014PhRvD..90j3527B}
\begin{eqnarray}
\dot{\overline{\rho}}_0&=& - 3\mathcal{H}\overline{\rho}_0 - a\Gamma \overline{\rho}_0 \, , \label{eq:rho0}\\
\dot{\overline{\rho}}_1&=& - 4\mathcal{H}\overline{\rho}_1 + \epsilon a\Gamma \overline{\rho}_0 \, , \label{eq:rho1}\\
\dot{\overline{\rho}}_2 &=& - 3(1+w_2)\mathcal{H}\overline{\rho}_2 + (1-\epsilon) a \Gamma \overline{\rho}_0 \, ,\label{eq:rho2}
\end{eqnarray}
where $\overline{\rho}_i$ is the background energy density of species $i$ and derivatives are again with respect to the conformal time $\eta$. The dynamical equation of state of the massive daughter particle $w_2(a)$ is 
\begin{equation}
w_2(a) = \frac{1}{3} \langle v_2^2(a) \rangle \, , 
\end{equation}
where $v_2$ is the speed of a massive daughter particle which was produced at an earlier time when $a=a_D$. By setting $\tilde{a} \equiv a_D/a$, we can write the average speed of the massive daughter as 
\begin{equation}
\langle v^2(\eta) \rangle = \int_{\eta_\star}^\eta v^2(\tilde{a}) \dot{n}_2{\rm d}\eta_D \Big/ \int_{\eta_\star}^\eta \dot{n}_2{\rm d}\eta_D \, , 
\end{equation}
where
\begin{equation}
v^2(\tilde{a}) = \frac{\tilde{a}^2 \beta^2_2}{1+\beta_2^2 \left[\tilde{a}^2-1\right]} \, ,
\end{equation}
$\beta_2=\epsilon/(1-\epsilon)$ is the speed of the massive daughter in units of the speed of light $c$ at the time of production, $\eta=\eta(a)$ is the conformal time that corresponds to scale factor $a$, and $\dot{n}_2 \equiv \mathrm{d}n_2/\mathrm{d}\eta_D$ is the time derivative of the massive daughter's number density. Finally, we use a constant $a_\star$ to define $\overline{\rho}_0(\eta=\eta_\star) = \rho_{c,0} \Omega_{\mathrm{cdm}}^{\mathrm{ini}}/a_\star^3$ with $\rho_{c,0}$ being the critical energy density today, $\eta_\star$ the conformal time for scale factor $a_\star$, and $\Omega_{\mathrm{cdm}}^{\mathrm{ini}}$ the initially assumed dark matter density \cite{2014JCAP...12..028A}. The initial conditions $\overline{\rho}_1(\eta=\eta_\star) = \overline{\rho}_2(\eta=\eta_\star)$ are set to be a small number that doesn't affect the early dynamics, and the initial population quickly becomes insignificant as more decays occur.

With the background evolution defined, we turn our attention to the perturbations of linear density, $\delta_i$; velocity, $\theta_i$; and shear, $\sigma_1$, as functions of the wavenumber $k$. The perturbations related to the parent particle are described by
\begin{equation}
\dot{\delta}_0 = -\frac{\dot{h}}{2} \, ,
\label{eq:delta_0}
\end{equation}
similar to standard cold dark matter with $\theta_0=0$. 
Perturbation evolution of the massless daughter particle  has  been extensively studied in \cite{2016JCAP...08..036P,2021PhRvD.103d3014C,2020arXiv200809615A,2021arXiv210212498A}, leading to the equations
\begin{eqnarray}
\dot{(\delta_1 r_1)} &=& -\frac{4}{3}r_1\theta_1 - \frac{2}{3}r_1\dot{h} + \dot{r}_1\delta_0 \label{eq:delta_1} \, ,\\
\frac{4}{3k}\dot{(\theta_1 r_1)} &=& \frac{k}{3}\delta_1 r_1 - \frac{4k}{3}r_1\sigma_1 \label{eq:theta_1}\, ,\\
2\dot{(\sigma_1 r_1)} &=& \frac{8}{15}\theta_1 r_1 +\frac{4}{15}r_1(\dot{h} + 6 \dot{\eta}) + \rm{h.o.}
\label{eq:sigma_1}
\end{eqnarray}
where $r_1=a^4\overline{\rho}_1/\rho_{c,0}$, $h$ and $\eta$ are the scalar metric perturbations. The higher order terms of the hierarchy of equations were terminated at the $\ell=17$ multipole, where $\delta$, $\theta$, and $\sigma$ correspond to $\ell=0$, 1, and 2 respectively (see \cite{2021arXiv210212498A} for a more detailed description).
On the other hand, the contribution of the massive daughter is a little more complicated. We adopt the warm dark matter fluid approximation scheme of \cite{2021arXiv210212498A} and calculate the continuity equation as 
\begin{eqnarray}
\dot{\delta}_2  &=& -3\mathcal{H}(c_{sg}^2 - w_2)\delta_2 - (1+w_2)\left(\theta_2 + \frac{\dot{h}}{2}\right) \nonumber \\
 &+& (1-\epsilon)a\Gamma \,  \frac{\overline{\rho}_0}{\overline{\rho}_2} \, (\delta_0 - \delta_2) \, ,
\label{eq:delta_2}
\end{eqnarray}
and the Euler equation as
\begin{eqnarray}
\dot{\theta}_2 & = & -\mathcal{H}(1 - 3c_{g}^2)\theta_2 + \frac{c^2_{sg}}{1+w_2}k^2\delta_2 -k^2\sigma_2 \nonumber \\
& -&(1-\epsilon)a\Gamma \frac{1+c_g^2}{1+w_2} \frac{\overline{\rho}_0}{\overline{\rho}_2}\theta_2 \, .
\label{eq:theta_2}
\end{eqnarray}
where $c_{sg}^2$ and $c_g^2$ are the synchronous gauge and adiabatic sound speeds respectively for the massive daughter and are defined as $c_{sg}^2 \equiv \delta P_2/\delta \rho_2$ and $c_g^2 \equiv \dot{\overline{P}}_2/ \dot{\overline{\rho}}_2$. The adiabatic sound speed can be written as
\begin{eqnarray}
c_g^2 &=& w_2\left[5-\frac{\mathfrak{p}_2}{\overline{P}_2}-\frac{\overline{\rho}_0}{\overline{\rho}_2}\frac{a\Gamma}{3 w_2 \mathcal{H}} \frac{\epsilon^2}{1-\epsilon} \right] \nonumber \\
& \times & \left[ 3(1+w_2)-\frac{\overline{\rho}_0}{\overline{\rho}_2}\frac{a \Gamma}{\mathcal{H}} (1-\epsilon)\right]^{-1} \, ,
\end{eqnarray}
where $\mathfrak{p}_2$ is known as the psuedo-pressure and it is a higher moment integral of background quantities \cite{2011JCAP...09..032L}.
For the synchronous sound speed, we follow the prescription in \cite{2021arXiv210212498A}
\begin{equation}
    c_{sg}^2(k) = c_g^2\left[ 1+ \frac{1-2\epsilon}{5} \sqrt{\frac{k}{k_{\mathrm{fs}}}}\right] \, ,
\end{equation}
where $k_{\mathrm{fs}}$ is the free-streaming length of the daughter particle
\begin{equation}
    k_{\mathrm{fs}} = \sqrt{\frac{3}{2}\frac{\mathcal{H}}{c_g}}\, .
\end{equation}

These two equations are sufficient to describe the contribution of the massive daughter particle to the perturbations. Based on our numerical tests, the contribution of the shear $\sigma_2$ or higher moments are negligible for the cold to slightly warm particles we are interested in here. We tested this up to $\epsilon<0.1$ and found at worst only percent level deviation. For more relativistic particles, this approximation breaks down, and one has to include higher moments.

\subsection{Effects of EDE and DDM on the CMB Power spectra}
\begin{figure}[t]
\centering
\includegraphics[width=0.95\columnwidth]{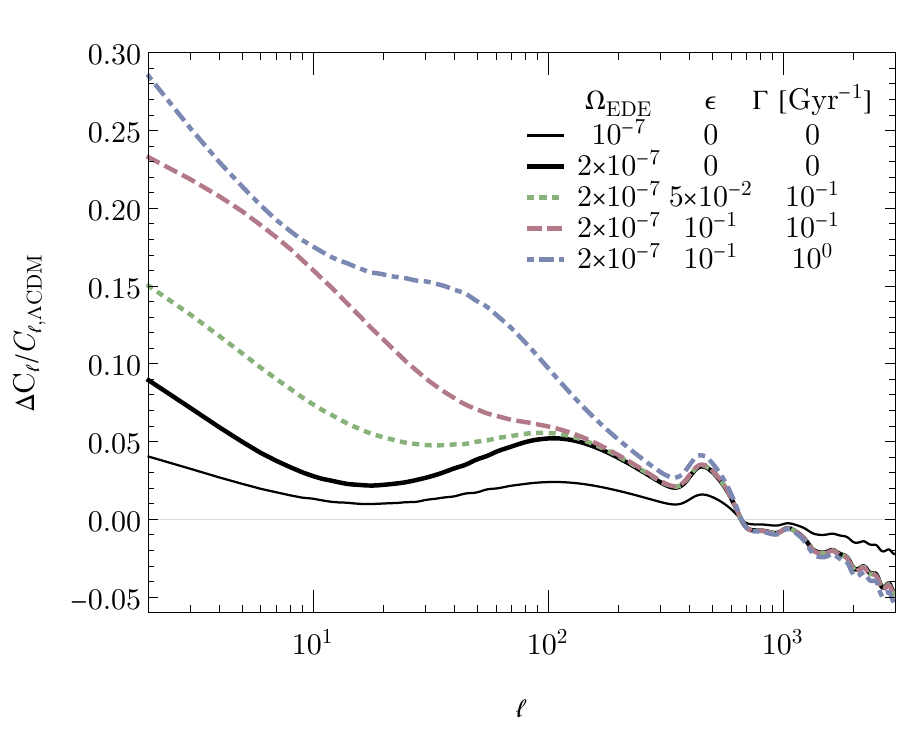}
\caption{Residual of the CMB temperature correlations for various models compared with the benchmark $\lcdm$ model. For EDE models, we choose $\log_{10}(a_\mathrm{EDE})=-3.7$ and $n=3$. The residuals are the two effects combined in an independent manner, resulting in enhancements on large scales and suppression on small scales. See the main text for details.}
\label{fig:cmb_diff}
\end{figure}

As expected from their individual effects, introducing both EDE and DDM results in alterations to the CMB power spectrum. For a simplified demonstration of these effects, we show in \Fig{fig:cmb_diff} the residuals of a comparison with a baseline $\lcdm$ model. We define our benchmark $\lcdm$ model with the following cosmological parameters: the peak scale parameter $100\theta_{\rm s} = 1.041783$, the baryon density today $\Omega_{\rm b} h^2 = 0.02238280$, the dark matter density today $\Omega_{\rm CDM} h^2 = 0.1201075$, the redshift of reionization $\tau_{\rm reio} = 0.05430842$, the matter power spectrum value $A_{\rm s} = 2.100549 \times 10^{-9}$, and the scalar tilt $n_{\rm s} = 0.9660499$ at the pivot scale $k=0.05$. These values are in agreement with Planck 2018 + lowE + lensing results~\cite{2020A&A...641A...6P}.

As shown in \Fig{fig:cmb_diff} by the bottom two solid curves, variations in $\Omega_{\mathrm{EDE}}$ causes an increase in the power spectrum correlations at large scales (small $\ell$) and a decrease at small scales (large $\ell$). These dependencies are the same for $n=3$ models as noted in earlier works, see \cite{2018PhRvD..98h3525P}.

Fixing the EDE model parameters, the largest alterations of DDM are due to the late-ISW effect as a result of time varying potential fields from the decaying particles and the transition to a $\Lambda$ dominated universe. With increasing $\Gamma$, this effect gets pushed to earlier times and increases on larger $\ell$. This could be observed in \Fig{fig:cmb_diff}: at high $\ell$'s, the dashed curves with non-zero $\Gamma$'s coincide with the black solid curve with the same $\Omega_{{\rm EDE}}$ but no decays. The two uniformed dashed lines (green and purple) deviate from the solid curve at the same $\ell$ ($\sim 100$) as they have the same $\Gamma$,
while the dot-dashed blue curve deviates at larger $\ell$ due to its larger value of $\Gamma$. In addition, for a given $\Gamma$, the relative magnitude of the alterations is controlled by $\epsilon$: the larger $\epsilon$ is, the bigger the deviation is.
The full dynamics is more complicated and is a combination of effects from varying both $\Gamma$ and $\epsilon$. All of these effects are detailed in prior works, see \cite{2021PhRvD.103d3014C, 2020arXiv200809615A,2021arXiv210212498A}.

From the residuals, we see that the general characteristics of the two individual components (EDE and DDM) are additive in our model. As it is presented in the discussion of \Fig{fig:cmb_diff}, the DDM's contributions add on top of the EDE ones. They behave independently of each other as they occur during different epochs. As discussed in the literature, EDE is consistent with a higher \H0, but it introduces a larger \S8. DDM cannot raise \H0 substantially, but it is able to force \S8 to lower values while remaining compatible with other observations. {\it These properties when combined together produce the desired effects of increasing \H0 and lowering \S8 simultaneously}. This is the main point of this paper, and we will now demonstrate it in detail in the following section.

\section{Results} 
\label{sec:results}
We perform a Markov Chain Monte Carlo (MCMC) analysis of the combined EDE and DDM cosmological model. We use \textsc{MontePython}~\cite{2018arXiv180407261B} and the Planck 2018 TTTEEE+low$l$+lowP+lensing data sets~\cite{2020A&A...641A...6P} in combination with other probes such as BAO (SDSS DR7\cite{2010MNRAS.404...60R}, 6FD\cite{2011MNRAS.416.3017B}, MGS\cite{2015MNRAS.449..835R}, BOSS DR12\cite{2017MNRAS.470.2617A}, eBOSS Ly-$\alpha$ combined correlations\cite{2019A&A...629A..85D,2019A&A...629A..86B}) and the Pantheon SNIa catalog\cite{2018ApJ...859..101S}. $\lcdm$ is modified with the addition of two variables, $\Gamma$ and $\epsilon$, for the DDM component and two other variables, $\Omega_{\rm EDE}$ and $a_{\rm EDE}$, for the EDE component. We use a modified version of \textsc{CLASS}\footnote{\url{http://class-code.net/}} ~\cite{2011JCAP...07..034B} to calculate the cosmological evolution and CMB anisotropies. We use the shooting method described in \cite{2014JCAP...12..028A} to compute the present-day dark matter density.

The MCMC analysis is conducted with the following flat priors:
\begin{eqnarray} 
&10^{-3}<\Gamma/(\mathrm{km/s/Mpc})^{-1}<10^{2.5} \, , \nonumber\\
&10^{-5}<\epsilon<10^{-1} \, , \nonumber \\
&10^{-20}<\Omega_\mathrm{EDE}<5\times 10^{-6} \, , \nonumber\\
&-5<\log_{10}(a_\mathrm{EDE})<-3.1 \, . \nonumber
\end{eqnarray}
The lower limits for $\Gamma$, $\epsilon$, and $\Omega_\mathrm{EDE}$ were chosen due to limitations in our numerical implementation which requires a non-zero quantity. However, these values are consistent with zero up to the uncertainties' of the experiments included in the analysis. The upper boundaries on these quantities were chosen to prevent numerical instabilities as their effects become large and deviate significantly from $\Lambda$CDM. The prior bounds for $a_\mathrm{EDE}$ were chosen such that the EDE transition is forced to occur during the matter dominated phase immediately preceding recombination, which is the typical transition time in EDE models.

\begin{figure*}[ht]
\centering
\includegraphics[width=1.95\columnwidth]{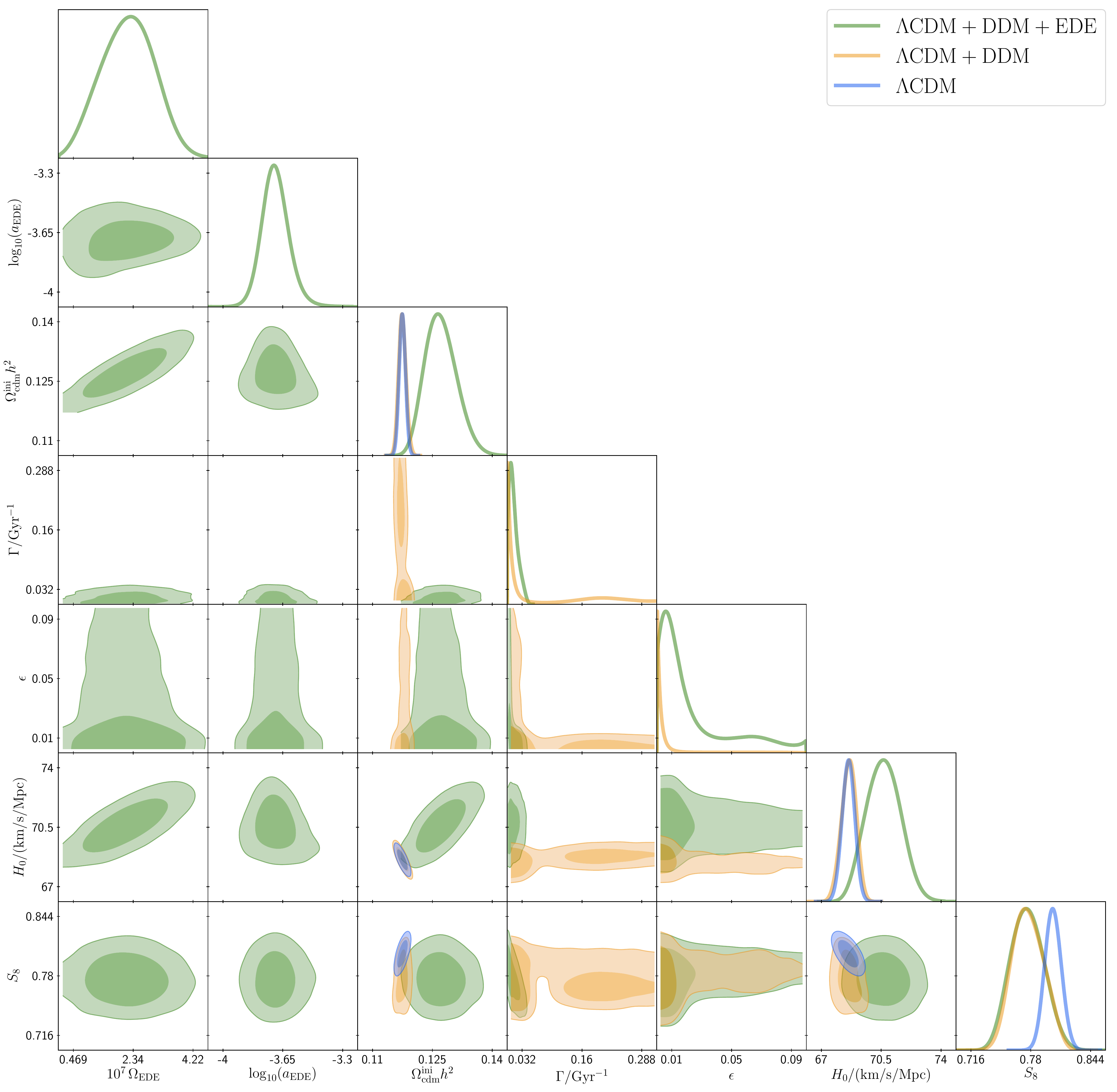}
\caption{Contour triangle plot of select model parameters demonstrating variations between the models: $\lcdm$ {(blue)}, $\lcdm$ + DDM {(orange)}, and $\lcdm$ + DDM + EDE {(green)} on the datasets Planck2018 + BAO + Pantheon + \S8 and \H0 priors. The major features of note are with the introduction of DDM, \S8 decreases while other parameters remain largely unchanged. With the introduction of EDE, some of the original parameters change substantially; in particular, both $\Omega_\mathrm{cdm}^\mathrm{ini}$ and \H0 prefer higher values. However, in contrast to EDE alone where \S8 increases, \S8 remains at its reduced value.
}
\label{fig:model_comp}
\end{figure*}

First, we study the effects of the addition of EDE and DDM to the benchmark $\lcdm$ when fitted on the entire data sets under the additional assumptions of an \H0 prior $73.2 \, \pm \, 1.3 \, \mathrm{km/s/Mpc}$ set by the SH0ES collaboration measurement \cite{2021ApJ...908L...6R} and an \S8 prior $0.766_{-0.014}^{+0.02}$ constructed from KIDS1000+BOSS+2dfLenS \cite{2020arXiv200809615A}. The results for select parameters are shown in \Fig{fig:model_comp}. The benchmark $\lcdm$ is shown in blue. As is well known, even with the use of the late universe priors, $\lcdm$ prefers a lower value of \H0 and a higher value of \S8, demonstrating both tensions.

The introduction of DDM is shown in orange on the same figure. Linear flat priors were assumed for both the decay rate $\Gamma$ and the energy fraction going into radiation $\epsilon$. In agreement with previous work \cite{2021PhRvD.103d3014C,2020arXiv200809615A,2021arXiv210212498A}, decays do not have a substantial effect on the value of \H0. \cite{2020arXiv200809615A,2021arXiv210212498A} also showed that DDM has a tendency to drive \S8 to lower values. Here, we observe this reduction with a median value of 0.78. We also observe a very interesting feature that is less apparent in previous studies which used a flat log prior for the DDM parameters \cite{2021PhRvD.103d3014C,2020arXiv200809615A,2021arXiv210212498A}. This feature is a splitting of the preferred parameter space into two distinct regions most apparent in the $\Gamma$ and \S8 plane. The two regions are separated by the DDM lifetime $\tau \sim 10 \; \mathrm{Gyr}$, corresponding to decays that have already occurred and those that are still ongoing. The two regions are joined at the top by a high $S_8$ region which corresponds to degeneracy between DDM and the benchmark $\lcdm$ model. Note that large values of $\epsilon$ are only obtained in the left region with smaller $\Gamma$, and the same low $\Gamma$ parameter space also allows for slightly higher $\Omega_\mathrm{cdm}^\mathrm{ini}h^2$. It is also important to mention that because the left and right $\Gamma$ regions are well separated, but of similar likelihood, the convergence between the two regions tends to cause chains to become stuck and can lead to one region being selected over the other on some runs.

We now turn our attention to the green contours for which the model includes both DDM and EDE. We again use flat priors for the parameters of the model and as expected \cite{2019PhRvL.122v1301P}, it raises the value of \H0 to 70.6 at the expense of an increased value of $\Omega_\mathrm{cdm}^\mathrm{ini}h^2$. The necessity of higher initial matter density for EDE to be able to resolve the \H0 tension, results in the selection of the low $\Gamma$ region. Without the decays, the increased matter would have also increased the value of \S8 \cite{2019PhRvL.122v1301P}, but with the decays, \S8 can be kept under control. The selection of the low $\Gamma$ region over the high $\Gamma$ region in order to reduce the matter contribution might seem counter-intuitive to initial expectations since fewer particles have decayed. However, this selection is due to $\epsilon$'s properties, particularly its influence on the lensing potential. Because more decays are needed in order to reduce the growth of structure in this model, for any given $\Gamma$, $\epsilon$ must increase. Changes to $\epsilon$ result in alterations to the lensing potential which are more substantial the early the decays occur. While $\epsilon$ must also increase in order to satisfy \S8, its time dependency is smaller. This combination results in preferring later decays and selecting the small $\Gamma$ parameter space. This is further confirmed by the higher median value of the one dimensional $\epsilon$ distribution compared to the one without EDE, as well as by the observation that in the bottleneck-like two dimensional distribution of $\epsilon$ against $\Omega_\mathrm{EDE}$, the highest values of $\epsilon$ are centered around the preferred EDE parameter values. A shape of this form also implies that a further increase of the EDE contribution would reduce $\epsilon$ in preference of a larger decay rate $\Gamma$, which is consistent with the slight positive correlation shown on the $\Gamma$ - $\Omega_\mathrm{EDE}$ panel. Most importantly, while DDM affects \S8, it doesn't interfere with the ability of EDE to increase the value of \H0 and thus both tensions could be relieved simultaneously.

Please note that while not that important for the low $\Gamma$ results, dark matter decays at redshifts $z \sim 2$ in the high $\Gamma$ region. This is well within the nonlinear regime, and for a complete analysis, the nonlinear effects of the decays should be taken into account; however, this is beyond the scope of this paper, and we leave it for future work.

\begin{figure}[ht]
\centering
\includegraphics[width=0.9\columnwidth, trim={0 0 6.32cm 6.22cm}, clip]{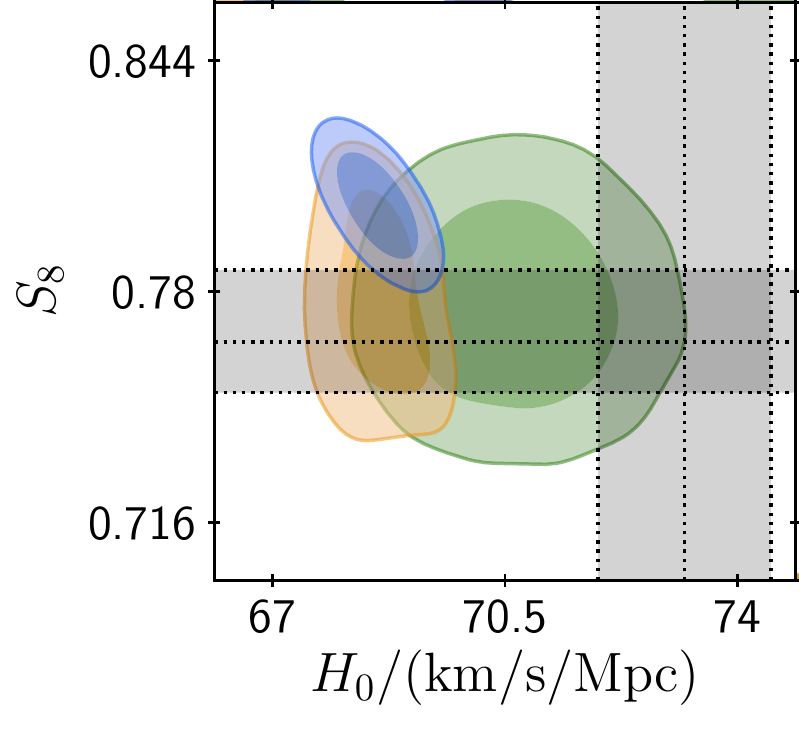}
\caption{A zoomed in version of the $S_8$-$H_0$ panel from \Fig{fig:model_comp} with $1\sigma$ shaded regions indicating the priors used. As would be expected, the priors pull the contours toward them as opposed to without (not shown). However, the magnitude is only significant for $S_8$ with the inclusion of DDM and $H_0$ with the inclusion of EDE.
}
\label{fig:model_comp_zoom}
\end{figure}

In \Fig{fig:model_comp_zoom}, we reproduce the \S8 - \H0 contours with the $1\sigma$ edges of the priors included. As would be expected and is evidenced with \H0 for EDE~\cite{2019PhRvL.122v1301P,2020PhRvD.102d3507H} and \S8 for DDM~\cite{2020arXiv200809615A,2021arXiv210212498A}, the priors drive the distributions towards the desired \H0 and \S8 values.

In order to determine the preference of $\lcdm$ + DDM + EDE over $\lcdm$, we calculate the Akaike Information Criterion ($\mathrm{AIC} = 2m-2\ln{\mathcal{L}_\mathrm{best}}$), where $m$ is the number of model parameters \cite{1974ITAC...19..716A}. This test is similar to the likelihood-ratio test with a penalty for additional degrees of freedom. We find a value of {$\Delta\mathrm{AIC}=-6.72$} which is just below a strong preference on the Jeffery's scale \cite{1939thpr.book.....J, 2013JCAP...08..036N, 2021arXiv210710291S}. Note that we found that $\lcdm$ + DDM and $\lcdm$ + EDE (not shown) have values of {$\Delta\mathrm{AIC}=-3.38$} and {-1.74}, which show only minor preferences.

\begin{figure*}[ht]
\centering
\includegraphics[width=1.95\columnwidth]{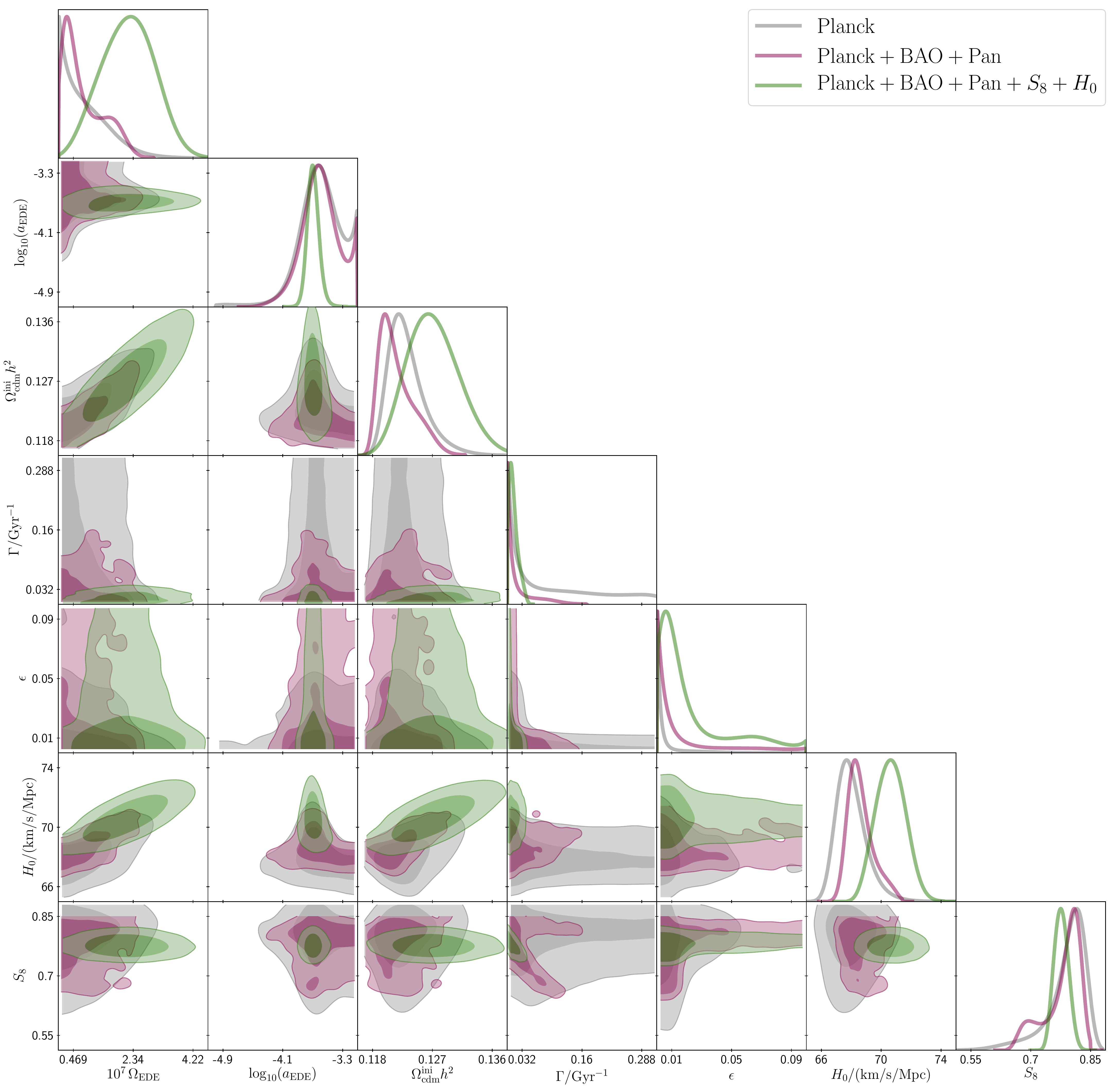}
\caption{Contour triangle plot of select model parameters of $\lcdm$ + EDE + DDM demonstrating variations between the contours when the datasets Planck18 {(grey)}, Planck18 + BAO + Pantheon {(purple)}, and Planck18 + BAO + Pantheon + \H0 and \S8 priors {(green)} are considered. The addition of datasets successively constrains the model with a preference for EDE and DDM once the priors have been included.
}
\label{fig:dataset_comp}
\end{figure*}

With the combination of EDE and DDM as a potential solution to both tensions, we examine the effects of different data sets on the posterior distributions. In \Fig{fig:dataset_comp}, we present the contours of the full model using Planck18 data alone \cite{2020A&A...641A...6P} (grey), Planck18 with the addition of BAO and Pantheon data \cite{2020A&A...641A...6P,2010MNRAS.404...60R, 2011MNRAS.416.3017B, 2015MNRAS.449..835R, 2017MNRAS.470.2617A,2019A&A...629A..85D, 2019A&A...629A..86B, 2018ApJ...859..101S} (purple), and finally, the full data sets with the the late universe \H0 and \S8 priors (green).

The first observation is that the Planck data {(grey)} alone does not significantly constrain the parameter space. This is expected because Planck by itself does not prefer DDM~\cite{2021PhRvD.103d3014C,2020arXiv200809615A,2021arXiv210212498A} and only shows mild preference for EDE~\cite{2019PhRvL.122v1301P,2020PhRvD.102d3507H}. However, the positive correlation in the $\Omega_\mathrm{EDE}$ - \H0 and the low \S8 regions in the decay parameters give mild credence to the model. As mentioned before, this increase in the \S8 posterior is the main motivation for this work. EDE has found success at alleviating the \H0 tension; however, this comes at the cost of an increase to the growth of structure. With the help of DDM, this weakness can be circumvented.

With the addition of the BAO and Pantheon datasets {(purple)}, the overall picture does not change significantly with the exception of $\Gamma$, which has a much reduced parameter space. It is particularly noteworthy in the $\Gamma$ - \S8 plane where we can see a splitting into two distinct regions. The upper branch is consistent with an \S8 value corresponding to that of $\lcdm$ and can be viewed as the degenerate overlap of the combined EDE + DDM model with $\lcdm$. The large range of allowed $\Gamma$ value in this branch, which also exists in the Planck {(grey)} contour, is a direct consequence of the anti-correlation of the decay parameters $\Gamma$ and $\epsilon$: the larger $\Gamma$ is, the smaller $\epsilon$ should be. The lower branch indicates a preferred region where a substantial amount of decays occur and this lowers \S8. Of special note (and expected from the degenerate behavior with $\lcdm$) is that the upper branch lies completely outside the \S8 prior while the lower branch lies strongly within. This indicates that the \S8 prior will select the lower branch. Note that as in the DDM case, the shape of the contour can cause the preferred region to differ between individual runs and increase the likelihood of chains becoming stuck. This is also one of the leading factors for the poor convergence in some of the purple contours.

For the last set of contours {(green)}, we add the \H0 and \S8 priors. Returning to the $\Gamma$ - \S8 plane, we see that only the lower branch is selected resulting in a preference away from $\lcdm$. However, while a preference is observed, the data sets used here are not sufficient enough to fully constrain $\Gamma$ or $\epsilon$, and we are only able to place upper bounds at the 68th percentile of {$\Gamma<1.72 \times 10^{-2}$~Gyr$^{-1}$ and $\epsilon<1.6\times10^{-2}$}. Another notable feature is a preference to have a non-zero $\Omega_\mathrm{EDE}$ as well as a fairly constrained time window for the EDE transition with mean values of {$\Omega_\mathrm{EDE} = 2.2 \times 10^{-7}$ and $\log_{10}(a_\mathrm{EDE}) = -3.7$}. This preference is driven by the \H0 prior. Finally, the last major change is a preference towards higher $\Omega_\mathrm{cdm}^\mathrm{ini}$. This is a direct consequence of EDE with the \H0 prior~\cite{2020PhRvD.102d3507H}. However, as we stated before, with EDE by itself, this leads to an increase in \S8; however, in combination with DDM, \S8 is controlled and even lowered to its preferred value. A summary of our results for the various model and data set combinations can be found in \Tab{tab:model_parameters}.

\begin{table*}[ht]
    \centering
    \begin{tabular}{|c|c|c|c|c|c|}
        \hline
        Model & ~$\lcdm$~ & ~$\lcdm$ + DDM~ & \multicolumn{3}{c|}{~$\lcdm$ + DDM + EDE~} \\ \hline
        Dataset & \multicolumn{3}{c|}{~Planck18 + BAO + Pantheon + \S8 + \H0~} & ~Planck18~ & ~Planck18 + BAO + Pantheon~ \\ \hline \hline
        $100~\Omega_{b}h^2$ & $2.260_{-0.013}^{+0.013}$ & $2.261_{-0.014}^{+0.014}$ & $2.275_{-0.020}^{+0.020}$ & $2.239_{-0.021}^{+0.018}$ & $2.251_{-0.019}^{+0.015}$ \\
        $100~\theta_{s}$ & $1.04211_{-0.00028}^{+0.00029}$ & $1.04211_{-0.00029}^{+0.00028}$ & $1.04138_{-0.00039}^{+0.00041}$ & $1.04163_{-0.00033}^{+0.00039}$ & $1.04177_{-0.00032}^{+0.00040}$ \\
        $\ln(10^{10}A_{s})$ & $3.048_{-0.015}^{+0.014}$ & $3.044_{-0.018}^{+0.016}$ & $3.064_{-0.017}^{+0.015}$ & $3.053_{-0.016}^{+0.016}$ & $3.052_{-0.017}^{+0.013}$ \\
        $n_{s}$ & $0.9716_{-0.0036}^{+0.0038}$ & $0.9719_{-0.0039}^{+0.0039}$ & $0.9825_{-0.0066}^{+0.0063}$ & $0.9679_{-0.0074}^{+0.0051}$ & $0.9710_{-0.0068}^{+0.0042}$ \\
        $\tau_\mathrm{reio}$ & $0.0581_{-0.0078}^{+0.0071}$ & $0.0567_{-0.0084}^{+0.0076}$ & $0.0594_{-0.0082}^{+0.0072}$ & $0.0559_{-0.0081}^{+0.0076}$ & $0.0570_{-0.0079}^{+0.0063}$ \\
        $\Omega_\mathrm{cdm}^\mathrm{ini} h^2$ & $0.11748_{-0.00083}^{+0.00080}$ & $0.1175_{-0.0010}^{+0.0009}$ & $0.1273_{-0.0042}^{+0.0036}$ & $0.1231_{-0.0035}^{+0.0016}$ & $0.1217_{-0.0032}^{+0.0015}$ \\
        $\Gamma/\mathrm{Gyr}^{-1}$ & $-$ & Unconstrained & $<0.0172$ & $<0.138$ & $<0.0256$ \\
        $\epsilon$ & $-$ & $<0.00476$ & $<0.0160$ & $<0.00527$ & $<0.0184$ \\
        $10^7\Omega_\mathrm{EDE}$ & $-$ & $-$ & $2.16_{-0.87}^{+0.81}$ & $<1.09$ & $<1.00$ \\
        $\mathrm{log}_{10}(a_\mathrm{EDE})$ & $-$ & $-$ & $-3.691_{-0.076}^{+0.064}$ & $-3.61_{-0.28}^{+0.26}$ & $-3.61_{-0.26}^{+0.24}$ \\ \hline
        $H_0/(\mathrm{km/s/Mpc})$ & $68.58_{-0.38}^{+0.38}$ & $68.60_{-0.42}^{+0.46}$ & $70.64_{-1.04}^{+0.96}$ & $68.00_{-1.19}^{+0.67}$ & $68.61_{-1.04}^{+0.54}$ \\
        $S_8$ & $0.8039_{-0.0092}^{+0.0090}$ & $0.777_{-0.019}^{+0.016}$ & $0.776_{-0.019}^{+0.017}$ & $0.792_{-0.021}^{+0.060}$ & $0.780_{-0.016}^{+0.056}$ \\ \hline
        $m$ & 28 & 30 & 32 & 31 & 32 \\
        AIC & 3938.70 & 3935.32 & 3931.98 & 2840.22 & 3926.72 \\
        $\Delta$AIC & $-$ & $-3.38$ & $-6.72$ & $-$ & $-$ \\ \hline
    \end{tabular}
    \caption{The mean with $1\sigma$ errors for the six principal $\lcdm$ model parameters plus the additional DDM and EDE parameters acquired from our analysis. Also shown are the derived parameters \H0 and \S8. To compare the various models, we present the Akaike Information Criterion $\mathrm{AIC} = 2m-2\ln{\mathcal{L}_\mathrm{best}}$ where $m$ is the number of model parameters. {$\Delta\mathrm{AIC}=-6.72$} indicates a preference for the combined model.\cite{1974ITAC...19..716A, 2013JCAP...08..036N} Note that $\Delta$AIC is only calculated for similar datasets.
    }
    \label{tab:model_parameters}
\end{table*}

\subsection{Linear vs log  priors}

\begin{figure*}[ht]
\centering
\includegraphics[width=1.95\columnwidth]{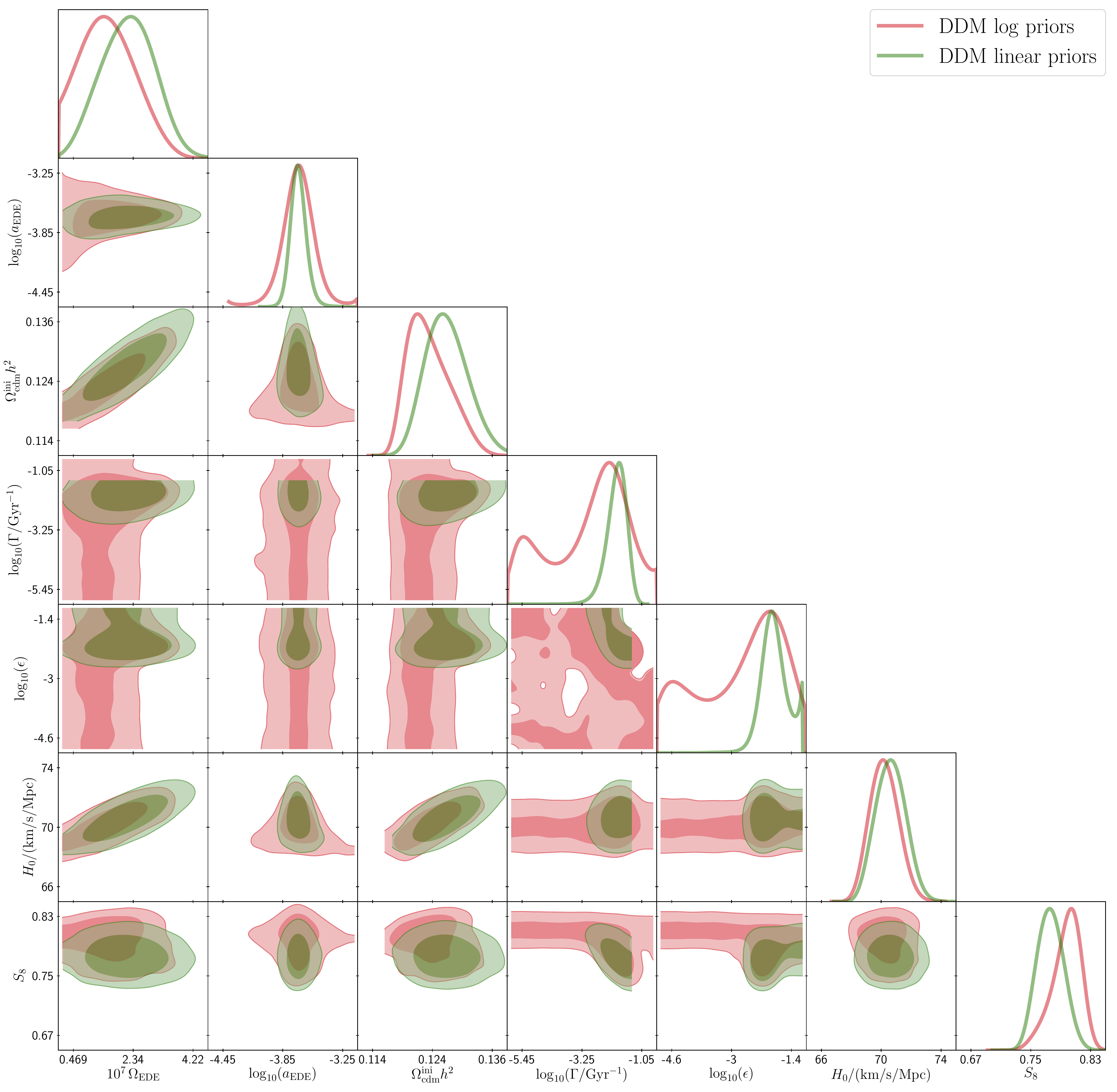}
\caption{Comparison between logarithmic {(red)} and linear {(green)} flat priors on the DDM parameters. Due to being degenerate with zero, the choice of lower bounds for log priors is arbitrary; however, this arbitrary choice increases the volume of the probed parameter space resulting in features being hidden or altered.
}
\label{fig:log_v_linear}
\end{figure*}

In prior works, the DDM parameter space was probed using flat-log priors~\cite{2021PhRvD.103d3014C, 2020arXiv200809615A, 2021arXiv210212498A}. Here, we give a brief investigation on the effect of having flat priors in log or linear space on the posterior distributions of the parameters of interest. A comparison between the two is shown in \Fig{fig:log_v_linear}.

While there are multiple features that can be pointed out, the most notable among the non-DDM parameters are \S8 - $\Omega_\mathrm{EDE}$/$\Omega_\mathrm{cdm}^\mathrm{ini}$ using the log priors {(red)}. Here, we observe two regions forming. The first coincides with the parameter space found in the combined model with linear priors {(green)}, which as seen in the \S8 - $\log_{10}{(\Gamma/\mathrm{Gyr})}$ or \S8 - $\log_{10}{(\epsilon)}$ planes, corresponds to non-zero DDM parameters and a reduced \S8. On the other hand, looking at the same planes, the log priors have a region extending to the left at constant $S_8$ distribution. This region coincides with an insignificant amount of decay energy transfer and is not present in the linear prior distribution. The left edge of this tail is obviously arbitrarily set by the lower edge of the priors. Because changes of the prior directly influence the volume of the entire preferred region, a lower log prior edge results in a reduction in the probability of residing in the parameter space of interest  where DDM can have an effect. Looking to the right side of the panel, both the log and linear priors produce a region that prefers lower \S8. However, a careful eye will notice that the log prior version is slightly smaller as a direct consequence of the volume effect mentioned earlier.

In other words, when we choose a log-prior for positive definite quantities, we run the risk of artificially increasing the volume of the parameter space by our choice of prior bounds. This occurs when the parameter does not have a well defined non-zero region and the posterior can only be described by a one-tailed distribution~\cite{2019JCAP...05..025D}. This is precisely the situation for the DDM parameters discussed in this work analysed using the current data sets and is the reason for the large degeneracy region with $\lcdm$ when using log priors. In addition, use of the log priors can potentially hide interesting features since the chains of the MCMC analysis could spend a lot of time in the uninteresting regions and under sampling regions of interest. This effect is also why the $\Gamma$ - $\epsilon$ contour is poorly converged in the logarithmic prior case. All points for either small $\Gamma$ or $\epsilon$ are equally likely, leading to poor contours. Since the effect of the smaller DDM parameter values is trivial and we wish to better sample parameter space of interest, we made the choice of using the linear priors to report our results.

\section{Conclusion} \label{sec:conclusion}
In this work we have studied the effects of expanding $\lcdm$ with the addition of a combination of Early Dark Energy and Decaying Dark Matter. We have shown that with the addition of EDE in the early universe and that of DDM in the late universe, both the \H0 and \S8 tensions can be reduced to within the 95th percentile uncertainties with {$H_0 = 71 \pm 1$} (reducing the tension to {$1.6\sigma$}) and {$S_8 = 0.78 \pm 0.02$} (removing the tension with the difference being {$0.4 \sigma$}). Our results show a preference for EDE with {$\Omega_{\rm{EDE}} = 2.1_{-0.9}^{+0.8} \times 10^{-7}$ and $\log_{10}(a_{\rm{EDE}}) = -3.69_{-0.08}^{+0.06}$} while setting an upper limit for DDM with {$\Gamma < 0.017 \; \rm{Gyr^{-1}}$ and $\epsilon < 0.016$}. We find that the combined model is preferred over $\lcdm$ with {$\Delta \rm{AIC} = -6.72$}. 

Our results indicate a strong preference for EDE, that  is in agreement with previous studies \cite{2019PhRvL.122v1301P,2019arXiv191010739N,2019arXiv191111760S} and is not affected by the presence of DDM. On the other hand, the posterior distributions of $\Gamma$ and $\epsilon$ are both consistent with zero, but the effect of the decays is evident on the rest of the cosmological parameters and they are playing an important role in restoring the \S8 value. Their presence removes any dependence of \S8 on $\Omega_\mathrm{cdm}^\mathrm{ini}$ and ensures that the late universe measurements will be in agreement. The current available datasets don't allow to set a lower boundary limit due to the degeneracy with $\lcdm$ and thus our results motivate further investigation of this scenario through other means, such as by studying velocity distribution disruptions in galactic halos. 

Finally, our results point to a probable more general characteristic of the \H0 and \S8 problems. Trying to solve either problem on its own has as a side effect other cosmological parameters being disturbed. Thus the two problems have to be addressed together, with a modification in the early universe needed to increase the \H0 value, due to the intermediate anchor of the BAOs limiting any late universe solutions and one independent late universe modification to address the inevitable increase in $\OmegaM h^2$ that such an early universe solution could introduce. The EDE-DDM is such an example of paired modifications but it is not necessarily the most efficient and further exploration is necessary.  

\section{Acknowledgments}
We thank Manuel Buen-Abad, Isabelle Goldstein, Leah Jenks, and Michael Toomey, Jatan Buch, and Vivian Poulin. JF is supported by the DOE grant DE-SC-0010010 and the NASA grant 80NSSC18K1010. SMK was partially supported by NSF PHY-2014052. Part of this research was conducted using computational resources and services at the Center for Computation and Visualization, Brown University.

\bibliography{manuscript}
\bibliographystyle{utphys_my_edit}

\end{document}